\documentclass[a4paper,12pt]{article}
\usepackage{graphicx}

\newcommand{\bmat}{\left(\begin{array}}
\newcommand{\emat}{\end{array}\right)}

\def\beq{\begin{equation}}
\def\eeq{\end{equation}}
\def\beqa{\begin{eqnarray}}
\def\eeqa{\end{eqnarray}}

\def\-{\hphantom{-}}

\def\s2{\frac{1}{\sqrt2}}

\def\beq{\begin{equation}}
\def\eeq{\end{equation}}
\def\beqa{\begin{eqnarray}}
\def\eeqa{\end{eqnarray}}
\def\ba{\begin{array}}
\def\ea{\end{array}}

\def\IF{\relax{\rm I\kern-.18em F}}
\def\II{\relax{\rm I\kern-.18em I}}
\def\IP{\relax{\rm I\kern-.18em P}}
\def\IC{\relax\hbox{\kern.25em$\inbar\kern-.3em{\rm C}}}
\def\IR{\relax{\rm I\kern-.18em R}}

\def\cp{{\cal P}}

\def\Dsl{\,\raise.15ex\hbox{/}\mkern-13.5mu D} 
\def\IZ{Z\kern-.4em  Z}

 \def\cp#1{\relax\ifmmode {\IP\kern-2pt{}_{#1}}\else $\IP\kern-2pt{}_{#1}$\=fi}

\catcode`\@=11
\newdimen\@rotdimen
\newbox\@rotbox

\def\@vspec#1{\special{ps:#1}}
\def\@rotstart#1{\@vspec{gsave currentpoint currentpoint translate
   #1 neg exch neg exch translate}}
\def\@rotfinish{\@vspec{currentpoint grestore moveto}}
\def\@rotr#1{\@rotdimen=\ht#1\advance\@rotdimen by\dp#1%
   \hbox to\@rotdimen{\hskip\ht#1\vbox to\wd#1{\@rotstart{90 rotate}%
   \box#1\vss}\hss}\@rotfinish}
\def\@rotl#1{\@rotdimen=\ht#1\advance\@rotdimen by\dp#1%
   \hbox to\@rotdimen{\vbox to\wd#1{\vskip\wd#1\@rotstart{270 rotate}%
   \box#1\vss}\hss}\@rotfinish}%
\def\@rotu#1{\@rotdimen=\ht#1\advance\@rotdimen by\dp#1%
   \hbox to\wd#1{\hskip\wd#1\vbox to\@rotdimen{\vskip\@rotdimen
   \@rotstart{-1 dup scale}\box#1\vss}\hss}\@rotfinish}%
\def\@rotf#1{\hbox to\wd#1{\hskip\wd#1\@rotstart{-1 1 scale}%
   \box#1\hss}\@rotfinish}%
\def\rotate{\@ifnextchar[{\@rotate}{\@rotate[l]}}
\def\@rotate[#1]#2{\setbox\@rotbox=\hbox{#2}\@nameuse{@rot#1}\@rotbox}

\catcode`\@=12

\begin{document}

\makeatletter \@addtoreset{equation}{section} \makeatother
\renewcommand{\theequation}{\thesection.\arabic{equation}}
\pagestyle{empty}
\pagestyle{empty}
\rightline{\today}
\vspace{2.0cm}
\setcounter{footnote}{0}


\begin{center}
\LARGE{\bf  Strange Attractors  in \\  Dissipative Nambu Mechanics :\\ Classical
and Quantum Aspects  }
\\[3mm]
{\Large{ Minos Axenides$^{1}$ and Emmanuel Floratos$^{1,2}$}}
\\[6mm]
 \normalsize{\em $^1$ Institute of Nuclear Physics, N.C.S.R. Demokritos,
GR-15310, Athens, Greece}\\
\normalsize{\em $^2$  Department of Physics, Univ. of Athens,
GR-15771 Athens, Greece}\\[5mm]  axenides@inp.demokritos.gr; \ \ \   mflorato@phys.uoa.gr
\end{center}
\vspace{2.0mm}
\begin{center}
{\large {\bf Abstract}}
\end{center}

\vspace{5.0mm}
{\small We extend  the framework of Nambu-Hamiltonian Mechanics
to include dissipation in $ R^{3}$ phase space.
We demonstrate that it accommodates the phase space dynamics of low dimensional
dissipative systems such as the much studied Lorenz and
R\"{o}ssler Strange attractors, as well as the more recent constructions of Chen
and Leipnik-Newton.
The rotational, volume preserving part of the flow preserves in time a
family of two intersecting surfaces, the so called { \em Nambu Hamiltonians}.
They foliate the entire phase space and are, in turn, deformed in time by
Dissipation which represents their irrotational part of the flow.
It is given by the gradient of a
scalar function and is responsible for the emergence of the Strange Attractors.

Based on our recent work on Quantum Nambu Mechanics, we provide an explicit
quantization of the Lorenz attractor through the introduction of Non-commutative
phase space coordinates as Hermitian $ N \times N $ matrices in $ R^{3}$.
They satisfy the commutation relations induced by one of the two Nambu Hamiltonians,
the second one generating a unique time evolution.  Dissipation
is incorporated quantum mechanically
in a self-consistent way having the correct classical limit without
the introduction of external degrees of freedom.
Due to its volume phase space contraction it violates
the quantum commutation relations. We demonstrate that the Heisenberg-Nambu
evolution equations for the
Quantum Lorenz system give rise to an attracting ellipsoid in
the $3 N^{2}$ dimensional phase space.

\newpage

\setcounter{page}{1} \pagestyle{plain}
\renewcommand{\thefootnote}{\arabic{footnote}}
\setcounter{footnote}{0}
\tableofcontents
\section{Introduction}
Dissipative dynamical systems, with a low dimensional phase space, present
an important class of simple non-linear physical systems
with intrinsic complex behavior (homoclinic bifurcations, period doubling,
onset of chaos, turbulence,...), which generated
intense experimental, theoretical
and numerical work in the last few decades
\cite{one, two, turbA, turbB}.

In this work we shall be interested to study dissipative dynamical
systems with a 3-dimensional phase space from the perspective of Nambu-Hamiltonian
Mechanics(NHM) \cite{Nam,Tak}. It represents a generalization of
Classical Hamiltonian Mechanics, which is mostly appropriate for the study of
phase-space volume preserving flows( Liouville's theorem).
To study dissipative systems in $ R^{3} $
we generalize NHM by splitting the flow vector field into its rotational(solenoidal) and
 irrotational components, which we identify them  to be its integrable, non-dissipative
 as well as  dissipative parts respectively.

We apply this idea to the most famous examples of Lorenz\cite{Lor}
and R\"{o}ssler
\cite{Ross}
 chaotic attractors
 which represent the prototype models for the onset of turbulence.\cite{RT,Feig,MP,LM}.
 We present also an Euler Top deformation of the Lorenz system
\cite{RBodA, RBodB} which will be a reference system for the quantization
of the chaotic attractors.
   Following our recent work on the Quantization of NHM \cite{AF,AFN}
   we provide a Matrix model
   for N-interacting Lorenz ( or R\"{o}ssler) attractors. It replaces the classical
   three phase space coordinates by three $ N \times N $ Hermitian matrices and the
    evolution equations with a symmetrically ordered matrix system
   (Weyl Quantization)\cite{Tar}. As a consequence we obtain a chaotic dynamical system
   of $ 3N^{2} $ real dimensional phase space for any integer  N.
      Finally we demonstrate that, similarly to the Lorenz system,
      there is an attracting
      ellipsoid in the $3N^{2} $-dim phase space, which provides
     a compact hypersurface for the localization
      of the higher dimensional
      attractors.

The plan of our paper is as follows:

       In sect. 2 we introduce the framework of Dissipative Nambu-Hamiltonian
       Mechanics in 3-dim phase space $ R^{3}$.
     Its Dynamics decomposes into a Non-dissipative integrable sector, which is
      parametrized in terms of two intersecting surfaces, the Nambu
      Hamiltonians (or Clebsch-Monge potentials of hydrodynamics) defined
     at  each point of the phase space flow.
     One of them defines (together with the initial conditions) a
     two dimensional
     phase space embedded in $R^{3}$, while the other defines the
     time evolution of the
     system, on this curved 2d-phase space.
     Their intersection defines the trajectory of the system and the
     flow vector field.
      The dissipative part of the flow is the gradient of a scalar function.

    In sect. 3 we apply the general framework to  the
       Lorenz system. We isolate a conservative sector parametrized by
       two intersecting surfaces:  a circular  and a parabolic cylinder .
       They  account, amazingly, for the double scroll topology of the full " butterfly"
       attractor. As a consequence of  the linear character of dissipation the
       trajectories of the full system get organized around
       the trajectories of the exactly integrable part.

       The Nambu-Hamiltonian form of the integrable part, given
       also in terms of Poisson brackets in $R^{3}$,
       defines, on one of the surfaces, a two dimensional phase space
       (the parabolic cylinder) with a standard  Poisson structure, of a
       quartic  anharmonic oscillator of single or double potential well,
       depending on the initial conditions.

 We present also an Euler top deformation of the Lorenz system .
 It is generated by a conservative sector given by the
intersection of a cylinder with a paraboloid.

In sect. 4 we apply our method method of analysis to the well known case
       of the R\"{o}ssler attractor.
       In this case the non-linear dissipative part distorts completely the orbits
       of the non-dissipative sector given by the intersection of
       a cylinder with an helicoid.

In sect. 5 we describe the basic steps of our Quantization
       prescription for Dissipative Nambu-Hamiltonian Mechanics.

In sect. 6 we construct in detail the Matrix Model for
       N interacting
       attractors as an illustration of our Quantization method.
       Moreover we demonstrate the existence of an attracting ellipsoid
       in $ 3 N^{2}$ dim. phase space.

Finally in sect. 7 we conclude our work with the basic results
       and a discussion of the relevance of our proposal to the formal
       quantization problem of dissipation of chaotic systems\cite{qdiss,qcha}.

\section{Dissipative Systems in $R^{3}$ and the Dynamics of Intersecting Surfaces}
Nambu-Hamiltonian mechanics is a specific generalization of classical
Hamiltonian mechanics, where the invariance group of canonical
symplectic transformations of the
Hamiltonian evolution equations in 2n dimensional phase space is extended to the
more general
volume preserving transformation group $ SDiff({\cal M}) $ with an arbitrary phase space
manifold {\cal M} of any dimension $d=\mbox{dim}({\cal M}) $ .

In the present paper we will
work with the case of $ {\cal M}= R^{3} $, a three dimensional flat phase space
manifold. Nevertheless our results could be generalized to curved manifolds of any
dimension\cite{Tak,AF}.

Nambu-Hamiltonian mechanics of a particular dynamical system in $R^{3}$ is defined once
two scalar functions $H_{i}\in C^{\infty}(R^{3}), i=1,2 $, the generalized
Hamiltonians \cite{Nam,Tak} are provided.
The evolution equations are:
\beq
\dot{x}^{i} \ = \  \{ x^{i} , H_{1} , H_{2} \} \ \ \ \ \ \ \ \  i=1,2,3
\eeq
where the Nambu 3-bracket, a generalization of Poisson bracket in
Hamiltonian mechanics, is defined as
\beq
\{ f, g, h \} \ = \ \epsilon^{ijk} \partial^{i}f \partial^{j}g \partial^{k}h
\ \ \ \ \ \ i,j,k=1,2,3 \ \ \ \ \ \ \forall f,g,h \in C^{\infty}( R^{3})
\eeq

Any local coordinate transformation
\beq
x^{i} \ \rightarrow \ y^{i} \ = \ y^{i}(x) \ \ \ \ \ \ \ \ \ i=1,2,3
\eeq
which preserves the volume of phase space
\beq
\mbox{det} \left( \frac{\partial y^{i}}{\partial x^{j} } \right) \ = \ 1 \ \ \ \ \
 \forall \ x=(x^{1},x^{2},x^{3}) \in R^{3}
\eeq
leaves invariant the 3-bracket and therefor it is a symmetry of Nambu-Mechanics.
Except for the linearity and antisymmetry of the bracket with respect to all of
its arguments it also
satisfies an important identity, the so called "Fundamental identity"[FI] \cite{Tak}.
Let
us define the operator
\beq
{\cal L}_{(h_{1},h_{2})} \ f \ \ = \ \ \{ h_{1} \ , \ h_{2} \ , f \ \} , \ \ \ \ \
 \forall \ h_{1},h_{2},f  \in C^{\infty}(R^{3})
\eeq
Then it can be shown that $ {\cal L} $ satisfies the commutation relation
\beq
[{\cal L}_{(h_{1},h_{2})} \ , \ {\cal L}_{(h_{3},h_{4})} ]\ \ = \ \
{\cal L}_{\{h_{1},h_{2},h_{3}\},h_{4}}
\ \ + \ \ {\cal L}_{h_{3},\{ h_{1}, h_{2},h_{4} \}} \ \ \
 \forall \  h_{i}\in C^{\infty}(R^{3}), \ \  i=1,2,3,4
\eeq
In \cite{AF} it was shown that rel(2.6) are the commutation relations of the Lie
algebras of Volume preserving
diffeomorphism Group SDiff($R^{3})$.
The argument goes as follows:
First using the operator $ {\cal L}_{(H_{1},H_{2})}$ the evolution equations ,
rel (2.1), can be written as
\beq
\dot{x}^{i} \ = \ {\cal L }_{H_{1},H_{2}} \ x^{i} \ \ \ \ \ \ \ \ i=1,2,3
\eeq
Given any function $ f \in C^{\infty}(R^{3}) $ of the phase space coordinates we obtain
the Liouville eq.
\beq
\dot{f} \ = \ {\cal L}_{H_{1},H_{2}} \ f \ \ = \ \ \{ f , H_{1} , H_{2} \ \}
\eeq
This equation demonstrates that $H_{1},H_{2}$ are constants of the motion and the orbit is
given by the intersection of the surfaces $ H_{1}^{o}=H_{1}^{o} , H_{2}=H_{2}^{o} $
where
\beq
H_{1}^{o} \ = \ H_{1}(x_{o}^{i}) , \ \ \ H_{2}^{o} \ = \ H_{2}(x_{o}^{i})
\eeq
with $x^{i}_{o}$ specifying the initial conditions at $t=0$, $x^{i}_{o}=x^{i}(0)$.
The formal integration of rel.(2.9) is:
\beq
f(x^{i}(t)) \ \ = \ \ e^{ t {\cal L}_{(H_{1}^{o},H_{2}^{o})}} \ \ f(x^{i}_{o})
\eeq
The evolution eq.(2.1) has a flow vector field:
\beq
v^{i}(x) \ \ = \ \ \epsilon^{ijk}\partial^{j} H_{1} \partial^{k}H_{2} \ \ \ \
\ \ \ \ \ i,j,k=1,2,3
\eeq
which is volume preserving
\beq
\partial^{i} v ^{i} \ = \ 0
\eeq
The reverse is also true. Any flow vector field of zero divergence
$ v^{i} \in C^{\infty}(R^{3}) , \ \ i=1,2,3 $ can be represented at least
locally by two Clebsch-Monge potentials $H_{1},H_{2}$ \cite{AF,Poly,Cleb,Lamb}
\beq
v^{i} \ \ = \ \ \epsilon^{ijk}\partial^{j}H_{1} \partial^{k}H_{2}
\eeq
From the representation rel.(2.13) and  the commutation relation for the
Lie algebra
of SDiff($R^{3}$) :
\beq
[ \ {\cal L}_{u} , {\cal L}_{v} \ ] \ = \ {\cal L}_{w}
\eeq
where $ {\cal L}_{u} = u^{i}\partial^{i} $ (similarly for $(v,w)$
with $ \partial^{i}u^{i}=\partial^{i}v^{i}=0 $ and
\beq
w^{i} \ = \ u\cdot \partial v^{i} - v \cdot \partial u^{i}
\eeq
follows that $ \partial^{i} w^{i} = 0 $ and hence the commutation
relations (2.6) follow.
We will name the phase-space volume preserving flows "Non-dissipative"
while the non-conserving ones "Dissipative".

The general flow vector
fields will have regions $ D_{+} \subset R^{3} $ with
\beq
\partial^{i} v^{i}(x) > 0 \ \ \ \ x\in D_{+}
\eeq
where any volume element is expanded by the flow and regions
$D_{-} \subset  R^{3}$, $ D_{+} \cap D_{-} = 0 $ with
\beq
\partial^{i} v^{i}(x) \ < \ 0 \ \ \ \ \ \ \  \forall x\in D_{-}
\eeq
where the volume elements are contracted.

The standard local stability analysis of the flow, involves the determination
of the critical points $x^{i}_{o} $(equilibrium points) where
$ v^{i}(x_{o}) = 0 , \ \ i=1,2,3 $,  as well as  the examination of the eigenvalues
of the fluctuation matrix
($ \partial^{i}v^{j}),{i,j}=1,2,3$)  at the critical points which
characterize the phase space portrait of the flow.

Under variations
of external parameters (environment) the location and stability character
of the critical points vary. There arises possibilities of Hopf bifurcations,
transition to chaos, turbulence etc \cite{one,two,turbA,turbB,RT}.

In order to bring into play the Nambu-Hamiltonian Mechanics framework
we firstly  generalize it for general dissipative systems in $R^{3}$ phase space.
We will decompose any flow vector field into its
rotational and irrotational parts.
\beq
v^{i} \ = \ \epsilon^{ijk} \partial^{j}A^{k} + \partial^{i} D \ \ \ \ \ \ \ \ i,j,k=1,2,3
\eeq
with the vector potential  determined up to a local gauge transformation:
\beq
A^{i} \ \rightarrow \ A^{i} \ + \  \partial^{i}\Lambda  \ \ \ \ \
\ \ i=1,2,3
\eeq
and
\beq
\partial^{i}v^{i} \ = \ \nabla^{2} D
\eeq
Given $ \partial^{i}v^{i} $ and appropriate BC we can , in principle,
determine the dissipation function D.

In general D is determined up to harmonic functions on $R^{3}$.
Since the rotational component  is divergenceless, we shall call it
the { \em Non-dissipative} part of the flow (ND), $ \vec{v}_{ND} $
while the irrotational
component we will identify it with the Dissipative part(D), $ \vec{v}_{D}$.
Once
the vector potential $A^{i}$ has been determined we can choose the
Clebsch-Monge gauge where [CM]:
\beq
A^{i} \ \ = \ \ H_{1} \partial^{i} H_{2}
\eeq
utilizing two "generalized Hamiltonians" or equivalently
Clebsch-Monge flow potentials.
The flow equations take the form
\beq
\dot{x}^{i} \ = \ \{ x^{i} , H_{1} , H_{2} \} \ + \ \partial^{i} D \ \ \ \ \ \ \ \
\ i=1,2,3
\eeq
and in vector form
\beq
\dot{\vec{x}} \ \ = \ \ \vec{\nabla }H_{1} \times \vec{\nabla }H_{2}
+ \vec{\nabla }D
\eeq
We notice that given a pair of functions $H_{1}, H_{2}$ such that
\beq
\vec{v}_{ND} \ \ = \ \  \vec{\nabla }H_{1} \times \vec{\nabla }H_{2}
\eeq
any transformation of $H_{1},H_{2}$
\beq
H_{i} \rightarrow H_{i}^{\prime }(H_{1}, H_{2}) \ \ \ \ \ i=1,2
\eeq
with unit Jacobian
\beq
\mbox{det} \left( \frac{\partial H_{i}^{\prime }}{\partial H_{j}} \right) \ = \ 1
\eeq
gives also
\beq
\vec{v}_{ND} \ = \ \vec{\nabla }H_{1}^{\prime } \times \vec{\nabla }H_{2}
^{\prime }
\eeq

Although the geometrical power of the surfaces $H_{1},H_{2}$  in the
Nambu mechanics framework ($D=0$) is lost, still there are interesting
cases to consider.
First the generalized Hamiltonians are not conserved but we have
\beqa
\dot{H_{1}} \ &=& \  \vec{\nabla}\vec{D} \cdot \vec{\nabla} H_{1}
\nonumber \\
\dot{H_{2}} \ &=& \vec{\nabla}D \cdot \vec{\nabla}H_{2}
\eeqa
and for D we obtain the equations:
\beq
\dot{D} \ = \  ( \vec{\nabla}D )^{2} \ + \ \{ D , H_{1} , H_{2} \}
\eeq
if the relative orientation of the surfaces is fixed of positive sign,
we get $ \{ D , H_{1}, H_{2}\} > 0 $ and D increasing for all times. In general
this is not true.

We also see that if D is a surface orthogonal to either $ H_{1}$ or
$ H_{2} $ then the latter are conserved.
 If D is orthogonal to both surfaces,
then it must be parallel to $ \vec{\nabla}H_{1} \times \vec{\nabla}H_{2} $
and the orbit of the dynamical system (2.24) is again the intersection of
$H_{1}$ and $H_{2}$ surfaces which are both conserved.

The decomposition of a general phase space flow vector field in its
(ND) and (D) parts may help to understand qualitatively
the phase space portrait behavior of the full system. Indeed, in principle,
it is possible to solve explicitly for the phase-space
trajectories of the ND part
\beq
\dot{\vec{x}} \ = \ \vec{v}_{ND} \ = \ \vec{\nabla } \ H_{1} \times \vec{\nabla }
\ H_{2}
\eeq
using the two integrals of motion
\beq
H_{i}(\vec{x}) \ = \ H_{i}(\vec{x}_{o}) \ \ \ \ \ \ \  i=1,2
\eeq
By parametrizing their intersection we can specify the explicit time dependence
of the motion.

In ref.\cite{AF} we reduced the evolution equation of the form (2.32)
Nambu Mechanics in Hamiltonian-Poisson form as follows:
\beq
\dot{x}^{i} \ = \ \{ x^{i} , H_{1} , \}_{H_{2}}
\eeq
where the induced Poisson bracket
\beq
\{ f , g \}_{H_{2}} \ = \ \epsilon^{ijk}\partial^{i}f \partial^{j}g \partial^{k} H_{2}
\eeq
satisfies all the required properties like linearity, antisymmetry
and the Jacobi identity:
\beq
\{ \{ f, g \}_{H_{2}} , h \}_{H_{2}} + \{ \{ g, h \}_{H_{2}} , f \}_{H_{2}} \ + \
\{ \{ h, f \}_{H_{2}} , g \}_{H_{2}} \ = \ 0
\eeq
The Jacobi identity follows from the Fundamental Identity rel.(2.7).

Since $H_{2}$ is conserved it defines a fixed surface
$ {\Sigma}_{2}: H_{2}=H_{2}(\vec{x}_{o})$ embedded in $R^{3}$ which can
be considered as a two dimensional phase space. In certain cases we may find
explicit local or global parametrization of $H_{2}$ with two parameters,
$ (u , v) \in {\Delta} \subset R^{2} $:
\beq
x^{i} \ = \ x^{i} (u,v) \ \ \ \ \ \ \ (u,v) \in \Delta \ \ \ \ i=1,2,3
\eeq
such that $ ( x^{i} ) \in \Sigma_{2} $
\beq
H_{2}( x^{i}(u,v) ) \ = \ H_{2} (\vec{x}_{o})
\eeq
The orbits on the $\Sigma_{2}$ phase-space are determined by the Hamiltonian $ H_{1} $ :
\beq
\dot{x}^{i} \ \ = \ \ \{ x^{i} , H_{1} \}_{\Sigma_{2}} \ \ \ \ \ \ \ \
\ i=1,2,3
\eeq
Comparing the relations on $ \Sigma_{2} $
\beq
\dot{\vec{x}} \ \ = \ \ \vec{x}_{u} \dot{u} \ + \ \vec{x}_{v}\dot{v}
\eeq
and
\beq
\dot{\vec{x}} \ = \ \vec{\nabla } H_{1} \times \vec{\nabla} H_{2}
\eeq
we get:
\beq
\dot{u} \ \cdot \ ( \vec{x}_{u} \times \vec{x}_{v} ) \ = \ \vec{\nabla}H_{2}
\ \cdot \ \partial_{v} H_{1}
\eeq
\beq
\dot{v} \ \cdot \ ( \vec{x}_{u} \times \vec{x}_{v} ) \ = \ - \vec{\nabla}H_{2}
\ \cdot \ \partial_{u} H_{1}
\eeq
Since $ \vec{x_{u}} \times \vec{x_{v}} $ and $ \vec{\nabla} H_{2} $
are normal and collinear at every point of $ \Sigma_{2} $
using parameters $ u, v $ of the same orientation (outwards) we must have:
\beq
\vec{x}_{u} \times \vec{x}_{v} \ = \ \rho(u,v) \ \  \vec{\nabla}H_{2}
\eeq
for a positive function $ \rho(u,v)$ on $\Sigma_{2} $,
we thus get from rel.(2.40-2.42)
\beq
\dot{u} \ = \ \frac{1}{\rho (u,v)} \ \ \partial_{v} \ H_{1}
\eeq
\beq
\dot{v} \ = \ -\frac{1}{\rho (u,v)} \ \ \partial_{u} \ H_{1}
\eeq
Relations $(2.43-2.44)$ can be written as Hamilton-Poisson eqs on $ \Sigma_{2} $ :
\beq
\dot{u} \ = \ \{ u, H_{1} \}_{\Sigma_{2}}
\eeq
\beq
\dot{v} \ = \ \{ v , H_{1} \}_{\Sigma_{2}}
\eeq
with the Poisson bracket on $\Sigma_{2} $:
\beq
\{ f , g \}_{\Sigma_{2}} \ = \ \frac{1}{\rho(u,v)} \ ( \partial_{u} f \partial_{v} g \ -
\ \partial_{u} g \partial_{v} f )
\eeq
We shall use instead the level-set Morse functions $H_{2}$ for the phase space
surfaces, because it is not always easy to find appropriate surface parametrizations.

We observe that from eq.(2.33), the Poisson structure of the $H_{2}$ phase-space, is
described in the algebra of phase space
coordinates $ x^{i}$ :
\beq
\{ x^{i} , x^{j} \}_{H_{2}} \ \ = \ \ \epsilon^{ijk}\partial^{k}H_{2}
\eeq
which is preserved by the eqns. of motion of the ND part.
As an example of Nambu-Hamiltonian dynamics and the reduced Hamilton-Poisson
form, we will consider the Euler free Top \cite{Nam,AF}.

For the Euler free top there are two conserved quantities. The total
angular momentum squared:
\beq
H_{2} \ = \ \frac{1}{2} \ \left( l_{1}^{2} \ + \ l_{2}^{2} \ + \ l_{3}^{2} \right)
\eeq
and the energy:
\beq
H_{1} \ = \ \frac{1}{2} \ \left( \frac{l_{1}^{2}}{I_{1}} \ + \ \frac{l_{2}^{2}}{I_{2}} \
+ \ \frac{l_{3}^{2}}{I_{3}} \right)
\eeq
where $ (l_{i}),i=1,2,3 $ are the components of the angular momentum, in the body frame,
and $ I_{1} , i=1,2,3 $ are the eigenvalues of the moments of inertia
tensor. The Nambu-Hamilton eqns :
\beq
\dot{l_{i}} \ \ = \ \ \{l_{i} , H_{1} , H_{2} \} \ = \ \epsilon^{ijk} \partial^{j}H_{1} \partial^{k}H_{2}
\eeq
provide the correct eqs. of motion for the rigid body :
\beqa
\dot{l_{1}} \ &=& \ ( \frac{1}{I_{2}} - \frac{1}{I_{3}} ) l_{2}l_{3} \ \ \ \ \ \nonumber \\
\dot{l_{2}} \ &=& \ ( \frac{1}{I_{3}} - \frac{1}{I_{1}} ) l_{3}l_{1} \ \ \ \ \ \nonumber \\
\dot{l_{1}} \ &=& \ ( \frac{1}{I_{1}} - \frac{1}{I_{2}} ) l_{1}l_{2}
\eeqa
The 2d-phase space is determined by the initial conditions:
\beq
l_{1}^{2}(t) + l_{2}^{2}(t) + l_{3}^{2}(t) \ = \ 2H_{2} \ = \ l_{o}^{2} =
l_{1,o}^{2} + l_{2,o}^{2} + l_{3,o}^{2}
\eeq
which  is an $S^{2}$ sphere in $R^{3}$.

The induced Poisson structure of the coordinates in $R^{3}$ is:
\beq
\{ l_{i},l_{j} \}_{S^{2}} \ \ = \ \ \epsilon_{ijk}\partial_{k} H_{2} =
\epsilon_{ijk}l_{k} \ \ \ \ \ \ \ \ \ \ \ \ \ \ \ \ \ i,j,k=1,2,3
\eeq
the Poisson $ SO(3) $ algebras and eqs. (2.52)
can be written
\beq
\dot{l}_{i} \ \ = \ \  \{ l_{i} , H_{1} \}_{S^{2}} \ \ \ \ \ \ \ \ \ i=1,2,3
\eeq

In the following sections  we are going to present a detailed investigation of the
Lorenz and R\"{o}ssler attractors from the point of view of Dissipative Nambu-Hamiltonian Dynamics.
\section{ The Lorenz System and its Euler top Deformation}
\subsection{ Lorenz Attractor from Dissipative Nambu Dynamics .}

The Lorenz model was invented as a three Fourier mode truncation
(Gallerkin approximation \cite{turbA}
of the
basic eqs. for heat convection in fluids in Reyleigh-Benard type of experiments
\cite{Sal,DG,Tab}
The time evolution eqns. in  the space of three Fourier modes $ x,y,z $ which we
identify as phase-space are :
\beqa
\dot{x} \ &=& \ \sigma ( y - x )    \nonumber  \\
\dot{y} \ &=& \ x(r-z) - y        \nonumber   \\
\dot{z} \ &=& \ xy - bz
\eeqa
where $\sigma $ is the Prandtl number, r is the relative Reynolds number
and b the geometric aspect ratio .

The standard values
for $ \sigma $, b are  $ \sigma=10, b = \frac{8}{3} $ with r taking values
in $ 1 \leq r < \infty $. There are dramatic changes of the
system as r passes through various critical values which follow the
change of stability character of the three critical points of the
system $P_{1} $ : $ x=y=z=0 , P_{\pm }: x=y=\pm \sqrt{b(r-1)}, z=r-1 $

Lorenz discovered the non-periodic deterministic
chaotic orbit for the value $r=28 $, which is today identified as a
Strange Attractor with a Hausdorff dimension of ($ d=2.06 $)\cite{FOY}. Standard
reference for an exhaustive numerical investigation of the Lorenz system
is the book by Sparrow \cite{Spa}, although there is an extended list
of numerical and analytic work on the system. For a qualitative discussion
of the physics we refer to \cite{Lor,MP,LM}.

We can have a first glimpse of the geometry of motion, for a volume element
in flow under eq.(3.1), if we cast the Lorenz system in Matrix form as follows:

\beq
\frac{d}{dt} \left( \begin{array}{c} x \\ y \\ z \end{array} \right)
\ = \ \left( \begin{array}{ccc}  -\sigma & \sigma & 0 \\ r & -1 & -x \\
0 & x & -b \end{array} \right) \left( \begin{array}{c} x \\ y \\ z
\end{array} \right) \ \equiv \  A \left( \begin{array}{c} x \\ y \\ z
\end{array} \right)
\eeq
and decompose the matrix A into its symmetric and antisymmetric parts
$ A= A_{s} + A_{as}$ :
\beq
A_{s} \ = \ \left( \begin{array}{ccc} -\sigma & \frac{\sigma + r}{2} &
0 \\ \frac{\sigma + r}{2} & -1 & 0 \\ 0 & 0 & -b \end{array} \right)
\eeq
\beq
A_{as} \ = \ \left( \begin{array}{ccc} 0 & \frac{\sigma - r}{2} & 0 \\
\frac{r - \sigma}{2} & 0 & -x \\ 0 & x & 0 \end{array} \right)
\eeq
The matrix $ A_{s} $ executes an expansion along an axis $x^{\prime }$,
a contraction along the orthogonal  $ y^{\prime} $ axis,  as well as a contraction
along the z-axis. The antisymmetric matrix $A_{as}$ executes
a rotation around z-axis with a constant $\frac{r-\sigma}{2}$ angular velocity and a rotation
around x-axis with an x dependent angular velocity.

The
axes $ x^{\prime} , y^{\prime} $ are rotated fixed axes which we find
through a diagonalization of $A_{s} $. We have thus the basic qualitative mechanism
for chaos through stretching, twisting and folding
\cite{turbB,RT,MP} .

 We turn our attention now to an important characteristic
property of the Lorenz attractor system, i.e. its "Orbit localizability"  within an
attracting(trapping) ellipsoid.\cite{Lor}

Consider the ellipsoid in $R^{3}$:
\beq
S_{1}\ = \ r x^{2} + \sigma y^{2} + \sigma(z -2 r)^{2}
\eeq
Along the orbits we have
\beq
\frac{d S_{1}}{d t} \ = \  \dot{\vec{r}} \cdot \vec{\nabla}S_{1} \ = \
-2 \sigma \ [ r x^{2} \ + \ y^{2} + \ b(z-r)^{2} \ - \ b r^{2} ]
\eeq
If we consider fixed values of $S_{1}$ such that the corresponding ellipsoid
$ S_{1}= \mbox{constant} $ lies entirely outside the fixed ellipsoid
\beq
S_{2} : \ \ \ \ rx^{2} + y^{2} \ + \ b(z-r)^{2} \ = \ b r^{2}
\eeq
then we get
\beq
\dot{\vec{r}} \cdot \vec{\nabla}S_{1} \ \ < \ \ 0
\eeq
for all orbits intersecting $S_{1}$, i.e. all of them are ingoing to the surface.
We observe that $S_{2}$ acts like an attractor region.
There
have been various attempts made to localize the Lorenz attractor, by convex surfaces,
in order to get information about Hausdorff dimensions \cite{FOY}
and other characteristics
\cite{DG,gattL}.

 The case of R\"{o}ssler attractor has been studied also
from this point of view\cite{gattR}.

 We will deal  with the issue of  {\em Localization} and the existence of an
 attracting ellipsoid in  our Matrix Model
 generalization of the Lorenz systems, in order to get attracting ellipsoids in higher
 dimensional phase spaces.

Concerning analytical methods we observe that the system can be explicitly
integrated out, through the use of Liouville operator.
For any observable $ \rho (x,y,z) $ on $ R^{3}$ we get the time
evolution on trajectories
\beq
\dot{\rho} \ = \ {\cal L} \cdot \rho
\eeq
with
\beq
{\cal L} \ = \ \sigma(y-x)\partial_{x} \ + \ [ x ( r - z ) - y ]
\partial_{x} \ + \ ( xy - bz )\partial_{z}
\eeq
The formal solution for the x-coordinates is
$ ({\cal L}_{o} = {\cal L}(x_{o}, y_{o}, z_{o}))$ :
\beq
x(t) \ = \ e^{t {\cal L}_{o}} x_{o}
\eeq
Similarly for y and z respectively.
Since the trajectories are smooth and bounded, we must have convergent
Taylor series in time\cite{Tab,SD,K}:
\beq
x(t) \ = \ \sum^{\infty }_{k=0} x_{k} \frac{t^{k}}{k!}
\eeq
with
\beq
x_{k} \ = \ ( \frac{d}{dt})^{k} x(t) \mid_{t=0} \ \ \ \ \ k=0,1,2,\cdots
\eeq
From rel(3.11) we get
\beq
x_{k} \ = \ {\cal L}^{k}_{o} \ x_{o} \ \ \ \ \ \ \ \ \ \ \  k=0,1,2,\cdots
\eeq
So we can get as many coefficients $x_{k}$ we like by simply looking at the linear in
the derivatives $ \partial_{x_{o}}, \partial_{y_{o}}, \partial_{z_{o}} $,
part of $ {\cal L}^{k}_{o} $.

We now proceed to describe the Lorenz system in the framework  of section 2.
The flow vector
field $ \vec{v} $ is analyzed into its dissipative and non-dissipative
parts as follows:
\beq
\vec{v}_{D} \ = \ ( -\sigma x , -y , -b z ) \ = \ \vec{\nabla } D
\eeq
with the "Dissipation" function
\beq
D\ = \ - \frac{1}{2} \ ( \sigma x^{2} \ + \ y^{2} \ + \ b z^{2})
\eeq
and
\beq
\vec{v}_{ND} \ = \ ( \sigma y, x(r-z), xy) \ = \ (0 , y , z-r) \times
(-x , 0 , \sigma )
\eeq
From rel (3.17) we get the two {\em Hamiltonians} or Clebsch-Monge potentials
$ H_{1}, H_{2} $
\beq
\vec{\nabla} H_{1} \times  \vec{\nabla} H_{2} \ = \ \vec{v}_{ND}
\eeq
or
\beq
H_{1} \ = \ \frac{1}{2} [ y^{2} + ( z- r )^{2} ]
\eeq
and
\beq
H_{2}\ = \ \sigma z \ - \ \frac{x^{2}}{2}
\eeq
The Lorenz system (3.1) can thus be written in the form ( $\vec{r}=(x,y,z) $):
\beq
\dot{\vec{r}} \ = \  \vec{\nabla} H_{1} \times  \vec{\nabla} H_{2}
\ + \ \vec{\nabla }D
\eeq
In the Non-Dissipative part(ND) of the dynamical system
\beq
\dot{\vec{r}} \ = \  \vec{\nabla} H_{1} \times  \vec{\nabla} H_{2}
\eeq
the Hamiltonians $ H_{1}, H_{2} $ are conserved and their intersection
defines the ND orbit.
Moreover if we get the reduced Poisson structure
(sect.2) from $ H_{2} $ we obtain the 2-dim phase space
$ \Sigma_{2}$  to be the family of parabolic cylinders
with symmetry axis the y-axis:
\beq
H_{2}=\mbox{constant} = H_{2}(\vec{r_{o}})
\eeq
 $ \Sigma_{2}$ is thus given by
 \beq
 z\ = \ z_{o} + \frac{x^{2} - x_{o}^{2}}{2\sigma}
 \eeq
 with $ x_{o},z_{o}$ the initial condition for $ x,z $.
 The induced Poisson algebra (rel. 2.33) is given by
 \beqa
 \{ x , y \}_{H_{2}} \ &=& \ \partial_{z}H_{2} \ = \sigma \nonumber \\
 \{ y , z \}_{H_{2}} \ &=& \ \partial_{x}H_{2} \ = -x \nonumber \\
 \{ z , x \}_{H_{2}} \ &=& 0
 \eeqa

The dynamics on the 2d-phase space $ \Sigma_{2} $ is given by $H_{1} $
\beqa
\dot{x} \ &=& \ \{ x , H_{1} \}_{H_{2}} \ \ \ \    \nonumber \\
\dot{y} \ &=& \  \ \{ y , H_{1} \}_{H_{2}}            \nonumber \\
\dot{z} \ &=& \ \  \{ z, H_{1} \}_{H_{2}}
\eeqa
and $H_{1}$ is an anharmonic oscillator Hamiltonian with $( x / \sigma, y )$
conjugate canonical variables.  Using rel.(3.19-3.20) we get on $ \Sigma_{2}$:
\beq
H_{1} \ = \ \frac{1}{2\sigma^{2}} \ [ y^{2} \ + \ \frac{1}{2} \ ( x^{2} \ - \ a^{2})^{2}]
\eeq
with
\beq
a^{2} \ = \ x_{o}^{2} \ - \ 2 \sigma ( z_{o} \ - \ r ) \ = \
- 2 H_{2} \ + \ 2 \sigma r
\eeq
where  $ \frac{1}{\sigma^{2}} $ plays the role of the mass.
Depending on the initial conditions we may have a single well
$(a^{2} \leq 0 , H_{2} \geq \sigma r )$ or a double well
potential ( $  a^{2}>0 , H_{2} < \sigma r $ ) respectively.
The trajectories , the intersections
of the two cylinders , $H_{1}$ and $H_{2}$ with orthogonal symmetry axes $(x,y)$ may
either have one lobe left/right or may be running from the right to the left lobe. This
is reminiscent of the topology structure of the orbits of the
Lorenz chaotic attractor.
\begin{figure}[thp]
\centering
\includegraphics[scale=0.4]{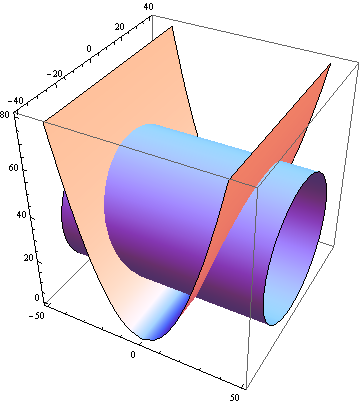}
\caption[]{Intersecting surfaces for Lorenz Attractor}
\label{Intersecting surfaces for Lorenz Attractor}
\end{figure}

In fig.1 we present the Lorenz Intersecting surfaces for
initial values $ x_{0}= z_{0}=0, y_{0}=1 $.
So the orbits of the non-dissipative system are the
seeds around which and through the action of dissipation, the actual attractor
is organized. In fig.2 we present such an example for orbits of the nondissipative
sector of the lorenz system superimposed with the full Lorenz attractor for the
same initial conditions and standard values of the parameters
( $ \sigma=10, r=28, b=8/3 $).

\begin{figure}[thp]
\centering
\includegraphics[scale=0.6]{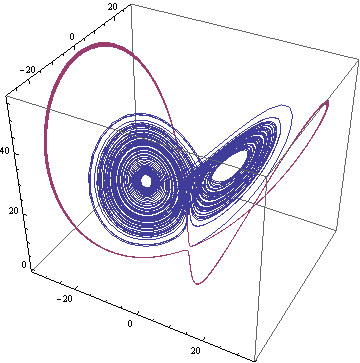}
\caption[]{Nondissipative Orbit and Lorenz Attractor}
\label{Nondissipative Orbit and Lorenz Attractor}
\end{figure}
We may observe directly the action of dissipation.
Similar observations have been
made by Haken, in a different physics context and Nevir-Blender \cite{Hak,NB}
without use of the induced
Poisson structure in $R^{3}$ rel.(3.25) and the ensuing interpretation \cite{AF}
(rel.3.26)
 Sparrow has
analyzed the integrable non-dissipative system, adding perturbative
dissipative terms \cite{Spa}.

The full Lorenz system does not conserve
$H_{1}, H_{2} $ and there is a random motion of the two surfaces against each other. Their intersection
is time varying.
In effect at every moment the system jumps from periodic to periodic orbit
of the non-dissipative sector.
Moreover the motion of the non-dissipative system around the two lobes,
either left or right, can now jump from time to time from one lobe to the other.

We would like to mention at this point ,another interesting feature of the Lorenz system.
Due to the linear character of the dissipative part of the Lorenz attractor ,we are able to
transform the full system to a volume preserving flow with time dependent
generalized Hamiltonians ( or equivalently intersecting surfaces).

This is done as
follows.
For a general   Nambu-Hamiltonian system with linear dissipation:
\beqa
\dot{x} \ &=& \ ( \vec{\nabla} H_{1} \times \vec{\nabla}H_{2} )_{1} \ -
\ \alpha x \ \ \ \ \nonumber \\
\dot{y} \ &=& \  ( \vec{\nabla} H_{1} \times \vec{\nabla}H_{2})_{2} \ - \ \beta y
\ \ \ \ \nonumber \\ \dot{z} \ &=& \ \vec{\nabla} H_{1} \times
\vec{\nabla}H_{2})_{3} \ - \ \gamma z
\eeqa
we define new variables (comoving frame) :
\beq
x= e^{-\alpha t} u \ \ , \ \ y = e^{-\beta t }v \ \ , \ \ z= e^{-\gamma t } w
\eeq
Using the gradient operators with respect to $ \vec{q} = (u,v,w) $ we can check
that the eqs. of motion (3.29) become
\beq
e^{-(\alpha + \beta + \gamma) t } \frac{d}{dt} ( u,v,w ) \ =
\ \vec{\nabla}_{\vec{q}} H_{1} \times \vec{\nabla}_{\vec{q}} H_{2}
\eeq
where
\beq
H_{i} \ = \ H_{i} ( e^{-\alpha t} u , e^{-\beta t}v , e^{-\gamma t }w ) \ \ \ \ \ \ i=1,2
\eeq
Eq.(3.31) implies that the $(u,v,w)$ phase space volume elements are conserved.
In effect, we have a new  time variable $ \tau \ = \ \frac{1}{\alpha+\beta+\gamma}
e^{(\alpha+\beta+\gamma)t}$
\beq
\frac{d}{dt} (u,v,w) \ = \ \vec{\nabla}_{\vec{q}} H_{1}
\times \vec{\nabla}_{\vec{q}} H_{2}
\eeq

The Lorenz attractor can be brought to the form of rel.(3.33). In terms
of our Hamiltonian reduction it is a Hamiltonian system with a time dependent
Hamiltonian $ H_{1}( e^{-\sigma t}u , e^{-t} v , e^{-b t} w )$ on a dynamical two
dimensional phase space which is defined by the surface
$ H_{2}(e^{-\sigma t} u , e^{- t} v , e^{-b t} w)=\mbox{constant} $ which
moves in $ R^{3} $. In the comoving coordinates $ u,v,w $ the evolution
of the system is volume preserving. A direct way to see the time dependence
of $ H_{1} , H_{2} $ and D, is through their time evolution eqs. 2.28-2.29.
For the Lorenz attractor system we have
\beqa
H_{1} &=& \frac{1}{2} \ \ ( y^{2} + (z - r)^{2}) \ \ \ \ \ \ \ \ \ \ \nonumber \\
H_{2} &=& \sigma z \ - \ \frac{x^{2}}{2} \ \ \ \ \ \ \ \ \ \nonumber \\
D &=& \ \ -\frac{1}{2} \ \ ( \sigma x^{2} \ + \ y^{2} \ + \ b z^{2} )
\eeqa
Their normal vectors are respectively:
\beqa
\vec{\nabla} H_{1} \ &=& \ ( 0 , y , z-r ) \ \ \ \ \ \ \ \ \ \nonumber \\
\vec{\nabla} H_{2} \ &=& \ ( -x , 0 , \sigma ) \ \ \ \ \ \ \ \nonumber \\
\vec{\nabla} D \ &=& \ ( - \sigma x , - y , - b z )
\eeqa
Finally we get:
\beqa
\dot{H}_{1} \ &=& \ -y^{2} - bz \ (z-r) \ \ \ \ \ \ \ \ \nonumber \\
\dot{H}_{2} \ &=& \ \sigma x^{2} - \sigma b z \ \ \ \ \ \ \nonumber \\
\dot{D} \ &=& \  \sigma^{2} x^{2} + y^{2} + b^{2} z^{2} - (\sigma + r) \ x y +
(1 - b) \  x y z
\eeqa
Since the Lorenz attractor is bounded in phase-space, the coordinates $ x, y,z $
have maxima
and minima, which implies that the surfaces $ H_{1}, H_{2}, S $ have maximum
and minimum  positions. So the Lorenz attractor is bounded
by the volume enclosed by the maximum and minimum positions of these three
surfaces.
Estimates of the maximum and minimum values of the functions $ H_{1},H_{2}$
have been determined using the constraint of the bounding Lorenz ellipsoid
as well as other surfaces.

Closing we would like to point out that
our analysis of the Lorenz system applies equally to similar
dissipative systems in $R^{3}$ such
as the Chen  \cite{Chen} and the Leipnik-Newton \cite{RBodA}
strange attractor constructions.
They all share the same structure with the Lorenz system. It involves
a nonlinear, nondissipative,  integrable part which admits an
intersecting surface Nambu parametrization superimposed to a linear dissipative
sector.
Last but not least it should be remarked that the $R^{3}$
geometrical picture presented above is directly generalizable to
dissipative systems with $R^{2n+1}$ phase space dynamics.
In complete analogy it is expected that
$2n$ intersecting surfaces (Nambu Hamiltonians) generate the integrable sector
of the system.

\subsection{A Strange Rigid Body Attractor as Dissipative Dynamics of Intersecting
Paraboloid with a Cylinder}
We notice, in passing, that we can obtain a dissipative Euler top deformation
of the Lorenz system, through the addition of terms rel.(2.55) :
\beq
\dot{l}^{i} \ = \ \{ l^{i} , H_{1} \}_{S^{2}} \ + \ \partial^{i} D \ \ \
\ \ \ \ \ \ i=1,2,3
\eeq
Indeed as we can see, the Lorenz  attractor system
\beqa
\dot{l}^{1} \ &=& \ \sigma ( l^{2} - l^{1} ) \ \ \ \ \ \  \nonumber  \\
\dot{l}^{2} \ &=& \ l^{1} (r - l^{3}) - l^{2} \ \ \ \ \ \  \nonumber  \\
\dot{l}^{3} \ &=& \ l^{1} l^{2} - b l^{3}
\eeqa
can be written as a dissipative Euler Top through the choice
$ I_{2}=I_{3}=I $ for any $ I > 0 $ and  $ I_{1} = \frac{I}{1+I}$ but only for
the special relation $ \sigma=r $ of its free parameters
$\sigma, r$. The corresponding Dissipation function D is given by:
\beq
D \ = \ -\sigma l^{1}l^{2} \ + \ \frac{1}{2} \ [ \sigma (l^{1})^{2} +
(l^{2})^{2} + (l^{3})^{2} ]
\eeq
The last condition $\sigma=r $, comes about from the constraint that we add to the Euler
top non-dissipative flow,  $ \vec{v}_{ND}= ( 0 , -l^{1}l^{3}, l^{1}l^{2})$, an
 irrotational (gradient) flow :
\beq
\vec{v}_{D}= ( \sigma(l^{2} - l^{1} ), r l^{1} - l^{2} , -b l^{3})
\eeq
with the condition,
\beq
\vec{\nabla} \times \vec{v}_{D} \ = \ (0 , 0 , r - \sigma ) \ = \ \vec{0}
\eeq
We performed some numerical experiments for the values of $ \sigma= r $.
We find that there is a chaotic attractor for values $ \sigma=r=(25,26,27) $
and $ b=8/3 $.

We introduce  now, to a 2-parameter Euler top deformation of the Lorenz system
\beqa
\dot{l}_{1} \ &=& \ \left( \frac{1}{I_{2}} \ - \ \frac{1}{I_{3}} \right)l_{2} l_{3} \ + \
\sigma ( l_{2} - l_{1}) \ \ \ \ \ \ \nonumber \\
\dot{l}_{2} \ &=& \ \left( \frac{1}{I_{3}} \ - \ \frac{1}{I_{1}} \right) l_{1} l_{3} \ + \
r  l_{1} - l_{2} \ \ \ \ \ \ \nonumber \\
\dot{l}_{3} \ &=& \ \left( \frac{1}{I_{1}} \ - \ \frac{1}{I_{2}} \right) l_{1} l_{2} \ - \
b l_{3}
\eeqa
where we have chosen $ I_{3} \geq I_{2} > I_{1} $. For more general
linear terms we obtain the Leipnik-Newton linear control system on the
motion of a Rigid Body where for appropriate values of the coefficients a
double Lorenz attractor is obtained
\cite{RBodA}.

Performing a similar decomposition in a conservative(non-dissipative)
 and dissipative sectors of
our system we get
\beq
\vec{v}_{ND} \ = \ \left( ( \frac{1}{I_{2}} - \frac{1}{I_{3}}) l_{2}l_{3} +
\sigma l_{2}, (\frac{1}{I_{3}} - \frac{1}{I_{1}} ) l_{1}l_{3} + r l_{1},
(\frac{1}{I_{1}}- \frac{1}{I_{2}}) l_{1}l_{2} \right)
\eeq
and
\beq
\vec{v}_{D} \ = \ ( -\sigma l_{1} , -l_{2}, -b l_{3} )
\eeq
we can find two Hamiltonians $H_{1}, H_{2}$ and the Dissipation function D:
\beqa
\vec{\nabla}H_{1} \times \vec{\nabla}H_{2} \ &=& \ \vec{v}_{ND} \nonumber \\
\vec{\nabla}D = - ( \sigma l_{1}, l_{2}, b l_{3})
\eeqa
In order to get the two characteristic lobes of the attractor already
at this level we notice that we can write  the non-dissipative
flow of rel.(3.43) as :
\beqa
\vec{v}_{ND} \ = \ ( 0 , \beta l_{2} , \gamma l_{3} -r ) \times ( -l_{1} , \delta l_{2} ,
\sigma^{\prime} )
\eeqa
where
\beq
\beta= \frac{1}{I_{1}} - \frac{1}{I_{2}} \ , \ \gamma=\frac{1}{I_{1}} - \frac{1}{I_{3}} \ , \
\delta = \frac{1 - I_{2}/I_{3}}{1 - I_{2}/I_{1}}
\eeq
and
\beq
\sigma^{\prime} \ = \ \frac{\sigma - \delta r }{\beta}
\eeq
From rel(3.46) we get,
\beq
H_{1} \ = \ \frac{1}{2} \ [ \beta l_{2}^{2} \ + \ ( \gamma l_{3} - r )^{2}]
\eeq
an elliptic cylinder along the $l_{1}$ axis and
\beq
H_{2}= \sigma^{\prime} l_{3} - \frac{l_{1}^{2}}{2} \ + \ \delta \frac{l_{2}^{2}}{2}
\eeq
a paraboloid with elliptic cross section.
The Dissipation function D  is identical to the case of the Lorenz attractor:
\beq
D \ = \ -\frac{1}{2} ( \sigma l_{1}^{2} + l_{2}^{2} + b l_{3}^{2} )
\eeq
From the chosen ordering $ I_{3} \geq I_{2} > I_{1}, $ we get $ \beta > 0,  \delta \leq 0   $
($=0 $ Lorenz case ) and   $ \sigma^{\prime} > 0 $.
The $ H_{2}$, an elliptic paraboloid is oriented towards the positive $l_{3}$ axis.
The intersection of the two surfaces $H_{1}, H_{2} $ accounts for the existence
of the two lobes of the attractor. Depending on the initial conditions
it is possible that these lobes are connected with trajectories of the
non-dissipative system. In fig.3 we present the intersecting surfaces for the Euler-
Lorenz nondissipative system for the initial values $ x_{0}=z_{0}=0, y_{0}=1 $ and
values of the parameters $ \sigma=10, r=28, b=8/3, I_{1}=1, I_{2}=2, I_{3}=3 $.
\begin{figure}[thp]
\centering
\includegraphics[scale=0.6]{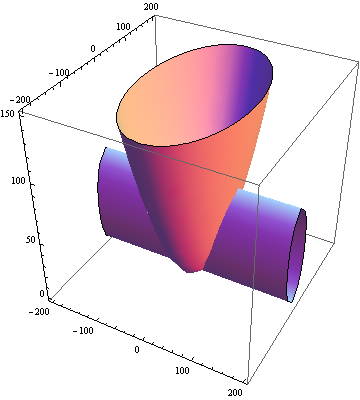}
\caption[]{Intersecting surfaces for Euler Top Attractor}
\label{Intersecting surfaces for Euler Top Attractor}
\end{figure}

A detailed analysis with numerical experiments will
be presented  in a forthcoming publication.
In fig.4 we present the nondissipative orbit for Euler-Lorenz system superimposed
with its full strange attractor for the same set of initial conditions and parameters
as in fig.3.
\begin{figure}[thp]
\centering
\includegraphics[scale=0.8]{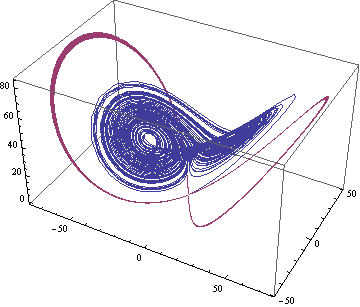}
\caption[]{Nondissipative orbit and Euler-Lorenz Attractor}
\label{Nondissipative orbit and Lorenz Attractor}
\end{figure}

\newpage

\section{ R\"{o}ssler Attractor from  Dissipative Dynamics of
a Cylinder Intersecting with an Helicoid}
R\"{o}ssler introduced a simpler than Lorenz's nonlinear ODE system with
a 3d- phase space, in order to study in more detail the characteristics of
chaos, which is motivated by simple chemical reactions \cite{Ross}.

The R\"{o}ssler system is given by the evolution eqns:
\beqa
\dot{x} \ &=& \ - y - z \nonumber \\
\dot{y} \ &=& \ x + a y \nonumber \\
\dot{z} \ &=& \ b + z(x-c)
\eeqa
with parameters $ a, b, c $ usually taking standard values
$ a=b=0.2, c=5.1 $ or
$ a=b=0.1, c=14 $ for the appearance of the chaotic attractor. The
two fixed points of the system are:
\beq
F_{\pm } : ( x_{\pm},y_{\pm}, z_{\pm} ) = ( \frac{c \pm \sqrt{c^{2}-4ab}}
{2} , - \frac{c \pm \sqrt{c^{2}-4ab}}{2a}, \frac{c\pm \sqrt{c^{2}-4ab}}{2a} )
\eeq
and depending on the parameters $a,b,c$ their stability character changes.
For the standard values  both are unstable saddle
points with the $ F_{-}$ having an unstable 2d manifold whereas the  $F_{+}$ an unstable
1d manifold. Studying the Poincare maps in the x-plane and the
bifurcation diagrams for $ y_{n+1}=f(y_{n}) $ return map, one discovers
the period doubling bifurcations to be the underlying mechanism for the
road to chaos\cite{ Feig}.

We turn now to the study of the R\"{o}ssler system as a Dissipative
Nambu-Hamiltonian dynamical system.

The key difference with the Lorenz attractor is that the dynamics of
the system is simpler. Chaos appears as random jumps outwards and inwards
the single lob attractor.

In order to get the three scalars,
the two generalized Hamiltonians $ H_{1} , H_{2} $ which are conserved and
characterize the non-dissipative part and D the  dissipation term :
\beq
\dot{\vec{r}}  \ = \ \vec{\nabla} H_{1} \times \vec{\nabla}H_{2} \ + \
\vec{\nabla} D
\eeq
we checked after some guess work, that we must subtract and add a new
term in the first equation. Indeed we find for the two parts
\beq
\vec{v}_{ND} \ = \ (-y - z - \frac{z^{2}}{2}, x , b )
\eeq
\beq
\vec{v}_{D} \ = \ ( \frac{z^{2}}{2}, a y , z (x-c) )
\eeq
where we impose the constraints
\beq
\vec{\nabla }\cdot \vec{v}_{ND} \ = 0
\eeq
and
\beq
\vec{\nabla} \times \vec{v}_{D} \ = \ 0
\eeq
So we must determine $ H_{1} , H_{2} $ and D such that
\beq
\vec{\nabla} H_{1} \times \vec{\nabla}H_{2}\ = \ \vec{v}_{ND}
\eeq
and
\beq
\vec{\nabla } D \ = \ \vec{v}_{D}
\eeq
For D we find easily
\beq
D \ = \  \frac{1}{2} \ [ a y^{2} + (x-c)z^{2}]
\eeq
To get $ H_{1}, H_{2} $ we must integrate first the Non-dissipative system:
\beqa
\dot{x} \ &=& \ -y -z - \frac{z^{2}}{2} \ \ \ \ \ \nonumber \\
\dot{y} \ &=& \ x \ \ \ \ \ \ \ \nonumber \\
\dot{z} \ &=& \ b
\eeqa
The general solution is :
\beq
x(t) \ = \ - b ( 1 + z(t)) \ + \ ( x_{o} + b(1+z_{o}) )\mbox{ cos(t)}
- ( y_{o} + z_{o} + \frac{z_{o}^{2}}{2} - b^{2} )\mbox{ sin (t)}
\eeq
\beq
y(t) \ = \ b^{2} - z(t) - \frac{z^{2}(t)}{2} +  ( y_{o} + z_{o} +
\frac{z^{2}_{o}}{2} - b^{2})\mbox{ cos(t)} \ + \ ( x_{o} + b + b z_{o} )\mbox{ sin (t)}
\eeq
\beq
z(t) \ = \ b t + z_{o}
\eeq
To uncover $ H_{1} , H_{2} $ we introduce the complex variable
\beq
w(t) \ = \ w_{1}(t) + i w_{2}(t)
\eeq
with
\beqa
w_{1}(t) \ &=& \ x(t) \ + \ b ( 1 + z(t) )  \nonumber \\
w_{2}(t) \ &=& \ y(t) + z(t) + \frac{z^{2}(t)}{2} - b^{2}
\eeqa
We obtain
\beq
w(t) \ = \ w_{o} \cdot e^{i t}
\eeq
with
\beq
w_{o} \ \equiv  \ w(t=0)
\eeq
We see that there are two constants of motion, the first one being :
\beq
\mid w(t) \mid  \ = \ \mid w_{o} \mid
\eeq
and we define correspondingly,
\beq
H_{1} \ = \ \frac{1}{2} \mid w(t) \mid ^{2} \ = \ \frac{1}{2} \
( x + b (1 + z) )^{2} \ +
\ \frac{1}{2} \ ( y + z + \frac{z^{2}}{2} -b^{2})^{2}
\eeq
The second integral of motion is obtained through the phase
\beq
w(t) \ = \ \mid  w_{o} \mid \cdot e^{\imath \varphi _{o}} \cdot e^{\imath t} \ =
\ \mid w_{o}\mid e^{\imath \varphi(t)}
\eeq
or from (4.14)
\beq
\varphi (t) \ - \ \frac{z}{b} \  = \ \varphi_{o} - \frac{z_{o}}{b}
\eeq
and we define appropriately the second constant surface $ H_{2} $
\beq
H_{2} \ = \ b \ \ \mbox{arctg} \frac{ y + z + \frac{z^{2}}{2} - b^{2}}
{1 + b(1+z)} \ \ - \ z
\eeq
We easily check that rel.(4.8) is satisfied:
\beq
\vec{\nabla} H_{1} \times \vec{\nabla}H_{2}\ = \ ( -y -z - \frac{z^{2}}{2} , x , b )
\eeq

The family of surfaces $ H_{1}$ and $ H_{2} $ are a quadratic deformation
of a cylinder and respectively a quadratic deformation of a right helicoid.
Their intersection is the trajectory (4.12-4.14).
\newpage
\section{ Quantization of Dissipative Nambu-Hamiltonian Dynamics }

The main motivation for us to introduce the framework  of Nambu-Hamilton
Dynamics into the Dissipative Dynamical systems, apart from the very useful
general global topological
information of the phase-space portrait, is the possibility provided
by this framework,for the formal quantization of chaotic attractors.

Indeed, based on our recent work \cite{AF},we are able to quantize systems,
for which the
two { \em Hamiltonians } $ H_{1} , H_{2} $ are general quadratic functions of the
phase space coordinates.
\beqa
H_{1} \ &=& \ \frac{1}{2} x^{i} M_{ij} x^{j} \  \nonumber \\
H_{2} \ &=& \ \frac{1}{2} x^{i} N_{ij} x^{j} \ \ \ \
\ \ \ \   i,j=1,2,3
\eeqa
where M,N are real symmetric matrices.

Since for the R\"{o}ssler attractor case
 we determined the Nambu surfaces $ H_{1} , H_{2} $ to be
 of the complex form rel.(4.20-4.23)
of higher degree than quadratic
(see 4.20-4.23 ),
we will proceed to work out in detail,
in the next and last section of this work, only the
quantization
of the Lorenz system .

In the present section we shall work out the
method of quantization for a general quadratic system.

At the end of the section
we will provide some practical way for a  Matrix Quantization of
general dissipative dynamical
systems in $R^{3} $(see also ref. \cite{Tar}).

Our strategy will be the following. Once the decomposition of the total
flow vector field between dissipative and non-dissipative parts has been established
we have the following representations :
\beq
\frac{ d \vec{x}}{d t} \ = \  \vec{\nabla} H_{1}
\times  \vec{\nabla}H_{2} \ +  \vec{\nabla} D
\eeq
The dissipative part $ \vec{v}_{D} $ is physically determined by either an
unspecified "external" environment or it corresponds to a phenomenological
modeling of our ignorance of some internal degrees of freedom.

We shall firstly quantize the non-dissipative system written in the form rel(3.26)
\beq
\dot x^{i} \ = \ \{ x^{i} , H_{1} \}_{H_{2}}
\eeq
In \cite{AF} we proposed the quantization of the Nambu 3-bracket and the induced
Poisson structure in $R^{3} $ as follows:
\beq
\{ x^{i} , x^{j} \}_{H_{2}} \ \equiv \epsilon^{ijk} \partial^{k} H_{2} \ \ \ \ \ \ \
\ \ \ \ \ \   i,j,k = 1,2,3
\eeq
Introducing a quantization(deformation) parameter $\hbar$  the phase-space
coordinates $ x^{i}$  go over to  operators for $  X^{i} $, satisfying
commutation relations of the form,
\beq
[ X^{i} , X^{j} ] \ = \ i \hbar \epsilon^{ijk} P^{k}(X) \ \ \ \ \ \ \
 \ \ \ \  i,j=1,2,3
\eeq
For polynomial Poisson brackets in (5.4) the corresponding
three operators $P^{k}, k=1,2,3 $ are respectively polynomials. They are
determined (non-uniquely)
by the following requirements

$\alpha$ )  The Jacobi Identity
\beq
[ X^{1}, P^{1} ] + [ X^{2}, P^{2} ] +  [ X^{3}, P^{3} ] \ = \ 0
\eeq
$\beta$ )  The Existence of a Classical Limit
\beq
   \lim_{\hbar \rightarrow 0} \ \ \ \ P^{k}(x) \ = \ \partial^{k} H_{2}
   \ \ \ \ \ \ \ \ \ \ \ \ \ \ k=1,2,3
\eeq
$\gamma$ ) The Existence of a Casimir
$ H_{2}(\hbar), \ \
( H_{2}(\hbar) \stackrel{\hbar \rightarrow 0}{\longrightarrow } H_{2}),$
\beq
[ X^{i} , H_{2}( \hbar ) ] \ = \ 0 \ \ \ \ \ \ \ \ \ \ \ \ i=1,2,3
\eeq
and the commutation relation (5.5) should permit the

$\delta$ ) {\em Diamond Property} of Unique Ordering of Monomials
\beq
(X^{i})^{m} \ ( X^{j} )^{n} ( X^{k} )^{l} \ \ \ \ \ \ i,j,k=1,2,3 \ \ \ \ \
\ \ \ \   m,n,l=1,2,\cdots
\eeq
into a sum of terms with given fixed order:
\beq
(X^{1})^{m_{1}} \  ( X^{2} )^{m_{2}} \ (X^{3})^{m_{3}} ,
\eeq
This last property generalizes the
Birkhoff-Witt theorem for polynomial enveloping algebras of Lie algebras.
The algebras (5.5) are called
Polynomial Lie Algebras \cite{SSV,Hop2}.  The case of quadratic $ H_{2} $
leads after quantization to classical Lie algebras.
So in case where:
\beq
H_{2} \ = \
\frac{1}{2} \ X^{i} N_{ij} X^{j}
\eeq
we get
\beq
[ X^{i} , X^{j} ] \ = \ i \hbar \epsilon^{ijk} N^{kl} X^{l} \ \ \ \ \ \ \ \ \ i,j,k,l = 1,2,3
\eeq
as well as the Heisenberg type of quantum evolution equations in $ R^{2} $ are
\beq
i \hbar \dot{X}^{i} \ = \ [ X^{i} , H_{1} ]
\eeq
where $ H_{1} $ is an ordered (say Weyl orderer) operator in $ X^{1},
X^{2},X^{3} $.
Each ordering determines a unique quantization procedure.

For example, for the quantization of the Euler top   we get
\beqa
H_{1} \ &=& \ \frac{1}{2} \left( \frac{ \hat{l}_{1}^{2}}{I_{1}} \ + \
 \frac{\hat{l}_{2}^{2}}{I_{2}} \ + \frac{ \hat{l}_{3}^{2}}{I_{3}} \right) \nonumber \\
 H_{2} \ &=& \  \frac{1}{2} \ \ ( \hat{l}_{1}^{2} + \hat{l}_{2}^{2} +
 \hat{l}_{3}^{2} )
 \eeqa
and for the eqs. of motion:
\beqa
\dot{\hat{l}}_{1} \ &=& \ \frac{1}{2} \ (\frac{1}{I_{2}} - \frac{1}{I_{3}} )
( \hat{l}_{2}\hat{l}_{3} + \hat{l}_{3} \hat{l}_{2} )
\nonumber \\
 \dot{\hat{l}}_{2} \ &=& \ \frac{1}{2} \ (\frac{1}{I_{3}} - \frac{1}{I_{1}} )
 ( \hat{l}_{3}\hat{l}_{1} + \hat{l}_{1} \hat{l}_{3} )
\nonumber \\
\dot{\hat{l}}_{3} \ &=& \ \frac{1}{2} \ (\frac{1}{I_{1}} - \frac{1}{I_{2}} )
( \hat{l}_{1}\hat{l}_{2} + \hat{l}_{2} \hat{l}_{1} )
\eeqa
with time preserved commutation relations
 $ [ \hat{l}_{i} , \hat{l}_{j} ] = i \hbar \epsilon_{ijk} \hat{l}_{k} $
where $\hat{l}_{i} $ are the standard angular momentum QM operators.
If we study the spin S system then the $\hat{l}_{i} s $ are $ (2s+1)\times (2s+1) $
Hermitian matrices.

We proceed, in what follows, to include the dissipative part in the quantization
procedure in such a way, that we recover the correct classical limit when the
quantization parameter $ \hbar $ goes to zero.
Quantum aspects of open dissipative systems have been the focus of
research in many areas of physics \cite{qdiss}.
A general operator quantization scheme for non-Hamiltonian systems was
recently developed by Tarasov \cite{Tar}.
In our present case  for a quadratic "Dissipation function"
D ( as it is the case for the Lorenz system) we have a unique way to recover the
classical limit, by choosing
\beq
D \ = \ \frac{1}{2} X^{i} S_{ij} X^{j}
\eeq
then
\beq
i \hbar  \dot{X}^{i} \ = \ [ X^{i} , H_{1} ] \ + \ i \hbar
S_{ij} X^{k}
\eeq
This form of quantization for quadratic $ H_{1} $, as it happens with
the Lorenz system

\beq
H_{1} \ = \ \frac{1}{2} X^{i} M_{ij} X^{j}
\eeq
gives the quantum evolution equation,
\beq
\dot{X}^{i} \ = \ A^{i}_{jk} \ \frac{1}{2} ( X^{j}X^{k} + X^{k} X^{j})
+ S^{ik} X^{k}
\eeq
where
\beq
A^{i}_{jk} \ = \ \epsilon^{ilm} M_{jl} N_{km} \ \ \ \ \ \ \ \ \
i,j,k,l,m=1,2,3
\eeq
with the correct classical limit.
The Weyl symmetric ordering leads to a general recipe for the
quantization of dynamical systems with polynomial flow vector fields,
in phase space coordinates. We simply utilize operator phase space coordinates,
symmetrize the right hand side of classical evolution eqs.
and forget about the
parameter $\hbar$. Comparison of different orderings can be done using
the commutation relations(5.12).

The algebra of coordinates  in the case of
the non-dissipative system is conserved in time
under the unique time evolution of $ H_{1} $, since the Casimir $H_{2}$ commutes with
any $H_{1} $. Including dissipation, $H_{2}$ and $H_{1}$ fail
to be conserved and the commutation relations rel(5.5) are not preserved in time.
In the next section we will consider this formal
quantization of the Lorenz attractor and we shall present the emerging Matrix
Lorenz Model.

Concerning the physical meaning of this formal quantization program, we would
like to stress the fact, that we are concerned here with quantum fluids,
physical systems
while macroscopic they keep their phase coherence(superfluids).
The various quantum fluid
motions are described by changes of action of the order of $ \hbar $.
The same holds true for the energy dissipation mechanism at the same space-time
scales. These conditions seem to hold in recently observed turbulence of
superfluid helium
[ $^{4}He, ^{3}He - B$  ]. Theoretical treatments of these experiments show
that probably even the classical fluid turbulence originates at the quantum
level \cite{DWSS}, although discussion is going
on about this point of view. We plan to come back on these interesting points
in our future work.
\section{  A  Matrix Attractor for the Non-commutative Lorenz System }
The Quantization method for Dissipative Nambu-Hamiltonian mechanics proposed
in the last section
corresponds to the passage from Hamilton-Poisson mechanics to Heisenberg's
quantum mechanical
evolution eqs. or matrix mechanics\cite{QMone}.
Although the basic duality principle of Quantum mechanics particle $\leftrightarrow$ wave,
is not directly transparent in Heisenberg's matrix mechanics the passage to
the Schrodinger's
picture, where the wave nature of quantum mechanical particles is automatic,
gets
easily established. In the case of Non-Dissipative
Nambu-Hamiltonian system in $R^{3}$ we generalized Hamilton-Poisson mechanics
by employing the induced Poisson structure on $ R^{3} $ by the 3-bracket :
\beq
\dot{x} \ = \ \{ x^{i}, H_{1}, H_{2} \} \ = \ \{ x^{i}, H_{1} \}_{H_{2}}
\eeq
The induced Poisson structure on $R^{3}$ is degenerate and reduces to a foliation
of non-degenerate Poisson structures, by the values of the level set Morse function
$ H_{2} = \mbox{constant} $.

Our proposal is a Matrix equivalent quantization in the line of Heisenberg's Quantum
Mechanics. The Lorenz system has both $H_{1}, H_{2} $ quadratic which is also the case
 for its Euler Top deformation as well as the Leipnik-Newton
system (sec. 3.2).
Indeed from section 3 for the ND Lorenz system
\beq
\dot{x}^{i} \ = \ \{ x^{i} , H_{1} , H_{2} \} \ \ \ \ \ \ \ \ \ \ i=1,2,3
\eeq
we have,
\beq
H_{1} \ = \  \frac{1}{2} \ [ y^{2} \ + \ ( z - r )^{2} \ ] \ \ \ \  \ \ \ H_{2} \ = \
 \sigma z - \frac{x^{2}}{2}.
\eeq
The $H_{2}$-phase space is an infinite parabolic cylinder with symmetry axis along
y-axis. Given the initial conditions $x_{o}, y_{o}, z_{o} $ it is given by
\beq
z \ = \ z_{o} \ + \ \frac{x^{2}-x_{o}^{2}}{2\sigma}
\eeq
with minimum at $x=0$,
\beq
z_{min} \ = \ z_{o} \ - \ \frac{x_{o}^{2}}{2\sigma}
\eeq
The corresponding Poisson algebra of coordinates in $R^{3}$ is:
\beq
\{ x , y \} \ = \ \sigma  \ \ \ \ \  \{ y , z \} \ = \ -x \ \ \ \ \ \{ z , x \} \ = \ 0
\eeq
Upon quantization we get for the $\hat{X},\hat{Y},\hat{Z}$ Hermitian operator
coordinates of the quantum mechanical phase -space
\beq
[ \hat{X} , \hat{Y} ] \ = \ \imath \hbar \sigma \ \ \ \ \ \  [ \hat{Y} , \hat{Z}  ] \
= \ -\imath \hbar \hat{X} \ \ \ \ \ \  [ \hat{Z} , \hat{X} ] \ = \ 0
\eeq

The $ \hat{X} , \hat{Z} $ operators can be simultaneously diagonalized and play the
role of two position "coordinates" while $ \hat{Y} $ is a conjugate "momentum"
operator to $ \hat{X} $.

Since $ \hat{H}_{2} $ is, by construction, the
Casimir of the algebra it must be real diagonal and for unitary irreducible
representations,  $\hat{H}_{2} $ must be a multiple of the identity operator. The
evolution Hamiltonian has been chosen to be $\hat{H}_{1}$
\beq
\hat{H}_{1} \ = \ \frac{1}{2} \ [ \hat{Y}^{2} \ + \ ( \hat{Z} - r )^{2} ]
\eeq
The quantum evolution equations for the Non-Dissipative Lorenz system are
\beqa
\imath \hbar \dot{\hat{X}} \ &=& \ [ \ \hat{X} , \hat{H}_{1} \ ]_{\hat{H}_{2}}    \nonumber \\
\imath \hbar \dot{\hat{Y}} \ &=& \ [ \ \hat{Y} , \hat{H}_{1} \ ]_{\hat{H}_{2}}    \nonumber \\
\imath \hbar \dot{\hat{Z}} \ &=& \ [ \ \hat{Z} , \hat{H}_{1} \ ]_{\hat{H}_{2}}
\eeqa
or using the algebra rel(6.7):
\beqa
\dot{\hat{X}} \ &=& \ \sigma \hat{Y} \ \ \ \ \ \ \ \ \ \ \ \ \ \nonumber  \\
\dot{\hat{Y}} \ &=& \ - \frac{\hat{X}\hat{Z}+ \hat{Z}\hat{X}}{2} \ + \ r \hat{X} \nonumber \\
\dot{\hat{Z}} \ &=& \  \frac{\hat{X}\hat{Y}+ \hat{Y}\hat{X}}{2}
\eeqa
We observe that  $\hat{H}_{1} $ along with the Casimir $ \hat{H}_{2} $ are conserved.

This has the consequence that for any time t the operators
$\hat{X}_{t},\hat{Y}_{t},\hat{Z}_{t}$  satisfy the same commutation relations(6.7).

The ND quantum Lorenz system is integrable as it is the classical one.
It is a quantum quartic  anharmonic oscillator. Indeed from rel.(6.4)
we get
\beqa
\hat{Z}(t) \ &=& \ \hat{Z}_{o} \ + \frac{(\hat{X}^{2}(t) -
\hat{X}^{2}_{o})}{2\sigma} \ = \
 \frac{X^{2}}{2\sigma} \ + \ \frac{\hat{H}_{2}}
{\sigma} \ \ \nonumber \\
 \ddot{\hat{X}} &=& \sigma \dot{\hat{Y}} \ =
 \ - \frac{\sigma}{2}
 \frac{XZ+ZX}{2} \ + \ r \sigma \hat{X}
 \eeqa
The Hamiltonian $\hat{H}_{1}$ becomes
\beq
\hat{H}_{1} \ = \ \frac{\hat{Y}^{2}}{2} \ + \ \frac{1}{8\sigma^{2}}
( \hat{X}^{2} \ - \ 2 ( \sigma r - \hat{H}_{2}) )^{2}
\eeq
We can deduce the eqn. for $\hat{X}$ directly from first two eqns. rel. (6.10- 6.11)
\beq
\ddot{\hat{X}} \ = \ -\frac{1}{2} \hat{X}^{3} \ + \ (\sigma r -
\hat{H}_{2}) \hat{X}
\eeq
With initial conditions
\beq
\hat{X}(t=0) \ = \ \hat{X}_{o} \, \ \ \ \ \dot{\hat{X}}(t=0) \ = \ \sigma \hat{Y}_{o}
\eeq
satisfying the algebra rel.(6.7)

Depending on the values of $\hat{H}_{2}$ we may have double
or single well quartic potential (repulsive or attractive linear term)
with a rich structure of physical states (discrete spectra, instantons,etc...).

We go back now to the full Dissipative quantum system.
According to our discussion in section 5 to have a natural classical
limit we shall apply the recipe of adding the dissipative terms (which are linear)
with the $ \imath \hbar $ prescription.

The question of going into  physical models of dissipation for the
Lorenz system, in the
framework of single mode lasers, has been discussed in \cite{ES,Grah}. There,
a specific form  is assumed, of quantum dissipation which is phenomenological.
The general question of dissipative quantum mechanics is discussed in several places
\cite{qdiss}.  Nevertheless there is no universally accepted way for improving
Quantum Mechanics including the environment into the Schrodinger or Heisenberg
picture(inclusion of the observer !).

According to sec. 5 we get
\beqa
\dot{\hat{X}}\ &=& \ \sigma (\hat{Y}- \hat{X})        \nonumber \\
\dot{\hat{Y}} \ &=& \ - \frac{\hat{X}\hat{Z}+ \hat{Z}\hat{X}}{2} \ + \
r \hat{X} - \hat{Y}      \nonumber \\
\dot{\hat{Z}} \ &=& \ \frac{\hat{X}\hat{Y}+ \hat{Y}\hat{X}}{2} \ - b\hat{Z}
\eeqa
Since the Casimir  $\hat{H}_{2}$  is not any more conserved, if we assume
the commutation relations (6.7) for the initial conditions
$\hat{X}_{o},\hat{Y}_{o},\hat{Z}_{o}$ they will not be preserved in time.
Indeed the role of dissipation is exactly the opposite of QM which assumes the
incompressibility of phase-space.

We shall now make a  motivated guess, that since
the classical dissipative Lorenz system has contracting volumes the same
occurs under certain conditions, to be determined, for the quantum case as well.
Motivated by this guess, which says that the effective quantum phase space
is compact( like in the Quantum Euler Top case) we proceed to study the
Operator valued(Quantum) Lorenz system of eqs. under the
assumption that $ \hat{X}, \hat{Y}, \hat{Z}$ are hermitian $ N \times
N $ matrices for appropriate $N=2,3,\dots  $. We make, in effect, the
proposal to study the Quantum mechanical system in the approximation
of a finite dimensional Matrix model (Finite Quantum Mechanics \cite{FQM1, FQM2}).

In section (3.2) we generalized the Lorenz system by adding some new terms,
 which makes the total system resembling as a deformation of the
free Euler Top.

 Moreover we observed
that the new system has similar attractor structure with the Lorenz one.
The new attractor is bigger in size but quite similar to the Lorenz one.
We propose to reverse the picture and consider the new system as
a dissipative deformation of the integrable Euler Top system (see the
Nambu-Hamilton representation in section 2) which breaks the rotational symmetry
down to a discrete $Z_{2}$ symmetry $ (x,y,z) \rightarrow (-x,-y,+z) $.

From this point of view
the quantization of these systems can be implemented, through the use of
N $\times$ N Hermitian matrices where $ N=2 s + 1 $, where s is the
spin of the Euler top ( see rel. 3.14-5.15 ). The quantum Lorenz system then is
a special case for $ I_{2}=I_{3}=I , I_{1}=\frac{I}{I+1} $ with added
dissipative terms.

It is easy to check that in this
case the multidimensional ellipsoid
\beq
S_{1}^{N} \ = \ tr \left( r \hat{X}^{2} \ + \ \sigma \hat{Y}^{2} \ + \
\sigma ( \hat{Z} - 2 r I )^{2} \right)
\eeq
appropriately chosen to be completely outside the ellipsoid
\beq
S_{2}^{k} \ = \ tr \left( r \hat{X}^{2} \ + \ \hat{Y}^{2} \ + \
b ( \hat{Z} - r I)^{2} \right)
\eeq
is in fact an attractor with any orbit passing through $S_{1}^{N}$ getting trapped.

The proposed Matrix Lorenz system for Hermitian $N \times N$ matrices
$\hat{X}, \hat{Y}, \hat{Z}$ rel.(6.16)
gives a non-linear ODE system with $3N^{2} $ equations. We observed that there is
a global $U(N)$  symmetry group in the system. For any $ U \in U(N)$, we have that
$ X\rightarrow UXU^{\dagger }, \  Y\rightarrow UYU^{\dagger }, \ Z\rightarrow
UZU^{\dagger } $ will be a solution if $ \hat{X}, \hat{Y}, \hat{Z}$ is one already.
As a result the invariant observables in the Matrix ODE v-system are traces of monomials
with the simplest to consider being of the form
\beq
tr( X) , tr( X^{2}) , tr(Y) , tr(Z), tr(XY) \cdots
\eeq
in terms of which the eigenvalues of $X, Y, Z $ matrices can be expressed.
Interesting plots appear to be the triples of eigenvalues $(\lambda_{i},
\mu_{i}, \nu_{i}), \ \ i=1,\cdots,N $ as functions of time :
\beqa
X(t) \ & \rightarrow & [ \lambda_{i}(t) ]_{i=1,\cdots,N} \nonumber \\Y(t) \ &
\rightarrow & [ \mu_{i}(t) ]_{i=1,\cdots,N} \nonumber \\
Z(t) \ & \rightarrow & [ \nu_{i}(t) ]_{i=1,\cdots,N }
\eeqa
Preliminary numerical experiments for $N=2,3 $ show Lorenz attractors with
more interactions between the two lobes. On the other hand the Matrix Lorenz
system (6.20) has an absorbing ellipsoid and so  each volume element shrinks to zero
asymptotically in time. The interesting question is, what is the Hausdorff
dimension of the Matrix attractor and under what condition
survives quantum mechanics or vice versa.

Concerning questions of interpretation of the matrix model we point out
that if  $X,Y,Z$ are
diagonal at time $ t=0$ they will remain so for any time. In this case we get
N decoupled Lorenz systems.
If we impose as initial conditions small off-diagonal
entries, then
we get weakly interacting Lorenz systems. On the contrary
large values for off-diagonal
terms are associated to strongly coupled Lorenz systems.
There is also a hierarchy of patterns for initial conditions which is preserved by
time evolution. We can have block-diagonal structures
with
\beq
N \times N \ = \ (N_{1} \times N_{1} ) \oplus (N_{2} \times N_{2} )
\oplus \cdots
 \oplus (N_{k} \times N_{k} )
\eeq
with small off-diagonal interactions in each block so there appears the possibility of having
patterns of self-similar structures as N grows. The non-commutative phase space structure
resembles the one of Matrix Theory\cite{BFSS}.

Extensions of
Lorenz systems has been also considered in the literature, with x,y complex numbers
and z real. Various Lorenz-like non-linear systems were
studied with weak dispersion and dissipation \cite{RBodB,GiGu}.

The interaction between
different Lorenz attractors with mismatch in their respective parameters
$ \sigma, r , b $ has attracted the interest of engineering(electronics-
telecommunication)biological and Chemistry scientific communities because of the phenomenon
of spontaneous appearance of either synchronization or  anti-synchronization in
amplitude or in phase
of the non-linear oscillator \cite{GPRD}.

Networks with nodes Lorenz non-linear oscillators
provide different universality classes of scale free networks\cite{CNets}.
We can extent in various ways our matrix model
transforming for example the parameters
$ \sigma, r , b $ into constant symmetric $ N \times N $ matrices (diagonal or not).

The question of introducing relativistic or not field theories by replacing
the harmonic oscillators of free fields (and the Feynmann  perturbative graphs)
by chaotic  non-linear oscillators has been proposed by C.Cvitanovic
in connection to the chaotic field theory and
with turbulence \cite{Cvi}.
Our Lorenz Matrix Model is a concrete idea exploiting this direction.

As far as the possible connection of the Lorenz Matrix model with the above growing
 interdisciplinary literature, we believe that it can serve as a new laboratory for
advancing theoretical issues on  chaos and turbulence both classical and quantum .

\section{ Conclusions-Open Problems}
The main result of our present work is the demonstration of Dissipative
Nambu-Hamiltonian mechanics
as the conceptual framework that underlies specifically strange chaotic attractors
both in their classical as well as quantum-noncommutative incarnation.
 Nambu's dissipative dynamics  of intersecting surfaces reproduced
 the familiar well studied attractor dynamics of Lorenz, R\"{o}ssler and
 Leipnik-Newton in a very
 intuitive manner accounting  of their gross topological aspects.
 Moreover we showed how it generates
 a novel deformation of the Lorenz attractor
 by embedding into it the Euler top dynamics. The detailed properties
 of our chaotic Quantum Euler Top Attractor will be
 investigated in the future.

 In all cases the dynamics of surfaces admits a
 noncommutative-quantum generalization through their fuzzification.
The general idea of non-commutativity of configuration spaces for fluids, fields
and of strings, provides a unifying framework for physics \cite{PolyB}.
 This, in  effect, led us to argue for the existence and construction ,
 in a systematic and consistent way, of a Matrix model for the quantum  Lorenz attractor.
It is a generalization of the well studied Lorenz
attractor to higher dimensional phase space.

This was essentially achieved through the decomposition of a general
chaotic flow into its rotational (non-dissipative) and irrotational (dissipative) parts.
It is conveniently described
by three scalar functions : the so called Clebsch-Monge potentials
(or Nambu Hamiltonians) $H_{1}, H_{2} $ as well as the Dissipation scalar
function D.
For each concrete physical system ,which is described by such (3d ODE's) flows
the identification of such a dissipative  or "environment" term is provided.
Generally this term, as was the case of the R\"{o}ssler attractor, is not an
purely irrotational vector field. It
encompasses phase space volume preserving components which we have to subtract
away and render it irrotational with the phase space portrait dynamics of the
trajectories as transparent as possible.

The non-dissipative term must exploit the basic topological characteristics of
the attractors via the geometrical intersection of
the constant surfaces $H_{1}, H_{2}$.

Given the constraints above
we could still have the freedom to determine the Dissipation  surface D
associated with the smallest possible energy loss.

In such a splitting, the non-dissipative and integrable sector
comprises the strongest term, an indicator of how close the entire system is to
being integrable or equivalently how big is the violation of its integrability
from pure dissipation.

Unexpectedly it uncovers gross topological features of the full dynamical
system (double lobe Butterfly for the Lorenz system and its Euler Top
deformation)
 or single lobe for
the R\"{o}ssler attractor.

The Quantum behavior of Strange attractors was  built systematically
through  { \em fuzzifying } the classical intersecting surfaces of the ND
sector .
We demonstrated this for the simplest case of the Lorenz system with a linear
dissipation
 together with the existence of an { \em attracting fuzzy ellipsoid}.
We constructed a
Matrix (hence noncommutative) Lorenz attractor
as a dynamical system of many coexisting attractors in 3-dim.
phase space, interacting with each other, through the off-diagonal terms
of the coordinate hermitian $N \times N $ matrices. The physical
picture that it implies is a concrete quantum mechanical model for the
transition to chaos and turbulence.

A decomposition into a real diagonal forms along with  the diagonalization of
unitary matrices of the phase-space coordinate matrices will make explicit
the role of the Quantum Mechanical phases and the wave nature of the system.
Synchronization phenomena in the amplitude or phase for the coexisting
attractors relate  equivalently to important issues concerning  decoherence
phenomena due to dissipation. Many Questions arise concerning Hausdorff
dimensions
mutual information and their scaling dependence on N deserve to be dealt with
in the future. Similarly the
behavior of time evolution of the eigenvalue distributions of the coordinate
matrices X,Y,Z are of importance for establishing the type of the generated
chaotic
ensembles.
Obviously we still at the beginning of our investigation and we hope that
this work
just opens an interesting set of questions.


\section{Acknowledgements}

We are grateful to J.Nicolis for sharing with us his insights on the theory
of chaotic strange attractors and their applications. For patiently guiding us
through their work
we are thankful to J.Gibbon, P.Nevir and V.Tarasov.
For discussions we thank C.~Bachas, I.~Bakas, C.~Kokorelis,
S.~Nicolis.
E.F. acknowledges partial support from the program { \em Capodistrias} at the
Univ. of Athens.


\end{document}